\documentclass[twocolumn,showpacs,preprintnumbers,amsmath,amssymb]{revtex4}

\usepackage{graphicx}
\usepackage{bm}

\begin{document}

\title{Metastable phases and "metastable" phase diagrams}

\author{V.V. Brazhkin}
\affiliation{Institute for High Pressure Physics, Russian Academy
of Sciences, Troitsk 142190, Moscow Region, Russia}

\date{\today}

\begin{abstract}
The work discusses specifics of phase transitions for metastable
states of substances. The objects of condensed media physics are
primarily equilibrium states of substances with metastable phases
viewed as an exception, while the overwhelming majority of organic
substances investigated in chemistry are metastable. It turns out
that at normal pressure many of simple molecular compounds based
on light elements (these include: most hydrocarbons; nitrogen
oxides, hydrates, and carbides; carbon oxide (CO); alcohols,
glycerin etc) are metastable substances too, i.e. they do not
match the Gibbs' free energy minimum for a given chemical
composition. At moderate temperatures and pressures, the phase
transitions for given metastable phases throughout the entire
experimentally accessible time range are reversible with the
equilibrium thermodynamics laws obeyed. At sufficiently high
pressures (1-10 GPa), most of molecular phases irreversibly
transform to more energy efficient polymerized phases, both stable
and metastable. These transformations are not consistent with the
equality of the Gibbs' free energies between the phases before and
after the transition, i.e. they are not phase transitions in
"classical" meaning. The resulting polymeric phases at normal
pressure can exist at temperatures above the melting one for the
initial metastable molecular phase. Striking examples of such
polymers are polyethylene and a polymerized modification of CO.
Many of energy-intermediate polymeric phases can apparently be
synthesized by the "classical" chemistry techniques at normal
pressure.
\end{abstract}

\maketitle

{\bf 1}. The phase transition effect falls equally within the area
of both physical and chemical studies. The notion of the "phase"
of was introduced by J.W. Gibbs \cite{1} as a condition of
substance with a certain "phase", meaning a set of pulses and
coordinates of the particles (atoms or molecules) that make up the
substance. At a later time, the meanings of the term "phase"
evolved so that currently the definitions of the phase of a
substance used in physics and chemistry literature are somewhat
different \cite{1,2,3,4,5,6}. In chemistry the thermodynamic
equilibrium condition is not mandatory for the phase. In physics
the thermodynamic non-equilibrium is allowed in principle but with
reservations of the kind as in \cite{2} "…sometimes the
non-equilibrium metastable state of a substance is also called the
"phase (metastable phase)". This seemingly minor distinction in
the definitions accepted in physics and chemistry has a profound
meaning. An incomplete understanding of the specifics of
metastable molecular organic and inorganic compounds often gives
rise to an erroneous interpretation of experimental data on phase
transitions in these systems. In the present paper, the focus of
our attention is the analysis of quasi-equilibrium and
non-equilibrium phase transformations in metastable phases.

    The name "metastable phase" is given to a non-equilibrium
    state of a substance whose properties change reversibly over
    a period of the experiment or observation. Strictly speaking,
    there are no metastable phases in classical thermodynamics
    since through the infinity of time the system should irreversibly
    relax to the equilibrium state. However, there are long-lived
    metastable solid phases whose life-time at normal conditions
    exceeds the time scale of the Universe, and their existence
    can not be neglected. The periods of investigation of phases
    in metastable state may vary within a wide range of $10^{-12}$ to $10^9$ s.
    A number of crystal metastable modifications, such as, for example,
    diamond, have the high pressure region of thermodynamic stability.
    Other metastable modifications, for instance, fullerite $C_{60}$ and
    amorphous solids, have no stability region at any pressure and
    temperature at all. In a certain range of the P,T-parameters,
    metastable phases can undergo reversible phase transformations
    with all the equilibrium thermodynamics laws obeyed
    (the transformations match the Gibbs' free energy equality for
    the corresponding phases) and with the reversible change of the
    structure of substances or even of their aggregate state
    (for example, during the melting and crystallization of white phosphorus).
    The phase mixture, too, can be metastable. Thus, J.W. Gibbs considered,
    by way of example, the mixture of hydrogen and oxygen which
    at moderate temperatures do not transform to the equilibrium
    phase $H_2O$ indefinitely long \cite{1}.

Metastable condition of phases is provided for by the existence of
the energy barrier on the way of the transformation of the system
into the low-laying energy states (see Fig.1). It is interesting
to note that the statistical physics still lacks effective ways of
checking the system for metastability. In other words, there is
not any possibility of determining whether the system is in the
local or the deepest minimum other than through the enumeration of
all the minimums and comparison of their depths. At reaching
certain pressures and temperatures, a high probability of
overcoming the corresponding energy barriers within the time of
the experiment (Fig.1) arises; in the process, the metastable
phases go beyond the quasi-equilibrium area of reversibility
(Fig.2) and relax, often through energy-intermediate states, to
thermodynamically equilibrium phases (Fig.2). Thus, on heating,
the amorphous modifications of substances crystallize; diamond, if
heated at normal pressure in the oxygen-free atmosphere at
1400-1700 K, transforms to graphite; the white phosphorus melt at
temperatures about 700 K crystallizes into violet-red phosphorus.
Fullerite $C_{60}$ on heating under pressure transforms to
graphite or diamond through the intermediate polymerized and
amorphous carbon states. Hydrogen-oxygen mixture on heating above
700 K begins reacting to form $H_2O$.

\begin{figure}
\includegraphics[width=7cm]{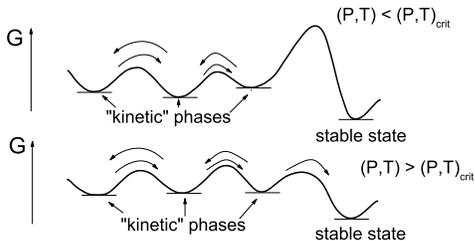}%
\caption{\label{fig:fig1} Energy minima in the configuration
space. At pressures and temperatures below critical
parameters metastable "kinetic" phases can transform into one
another but can not transit to the ground stable state with the
minimal Gibbs' free energy. At critical P,T-parameters the
irreversible transitions from the "kinetic" phases to the stable
modification occur for the experimental times. }
\end{figure}

All these transitions to more equilibrium states are fundamentally
irreversible kinetic processes, accompanied by the decrease in the
Gibbs' free energy and the corresponding heat release; they and
are not phase transformations in the strict sense. Accordingly,
for metastable phases one should examine a non-equilibrium kinetic
transformations diagram (transitional diagram) (Fig.2) rather than
an equilibrium phase diagram, with the part of the former
(P,T-reversibility region) being quasi-equilibrium. We emphasize
that the size of the quasi-equilibrium area of the phase diagram
depends on the period of observation or experiment. These are
commonly known considerations, but in physics the notion of
"metastable phases" is of somewhat exotic character since most of
simple substances studied by physicists refer to stable phases.

\begin{figure}
\includegraphics[width=7cm]{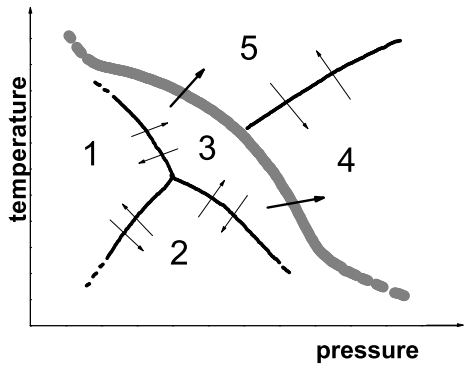}%
\caption{\label{fig:fig2} Transitional phase diagram of initially
metastable phases. The grey band corresponds to the irreversible
non-equilibrium relaxation of metastable phases to the stable
modifications and separates the quasi-equilibrium and equilibrium
P,T-regions. Inside these regions the transformations (1-2, 2-3,
1-3 and 4-5) are reversible. The transitions 3-4 and 3-5 are
irreversible. }
\end{figure}

{\bf 2}. When passing from elemental substances and simple
compounds to more complex compounds, especially of organic nature,
the ratio between stable and metastable phases radically changes.
The overwhelming majority of condensed phases of organic compounds
are metastable, i.e. they do not match the Gibbs' free energy
minimum for a given composition of elements \cite{7}. For yet more
complex systems of biological nature, metastable are 100 \% of
phases. In other words, as compounds become more complex the
proportion of thermodynamically equilibrium phases goes down, and
we materially live in the world of predominately metastable
phases. As we see, in chemistry the property of equilibrium fells
outside the concept of a "phase", and phase diagrams in fact are
often transitional diagrams (Fig.2) with the corresponding
quasi-equilibrium and irreversible relaxation regions. Thus, the
demarcation line between physics and chemistry of condensed media
in a certain sense lies in the degree of occurence of metastable
phases among the objects of research.

    So, the objects of research in physics are primarily stable
    phases of elemental substances and simple compounds, while
    chemistry is basically preoccupied with metastable phases
    of complex compounds. The difficulties and ambiguities usually
    arise in the border region when investigating the phase transitions
    in molecular substances based on light element compounds,
    including simple compounds in the system C-O-N-H. Because
    of relative simplicity of these compounds, the physicists
    look upon them as "their" objects, often being unsuspecting
    of the fact that the vast majority of the condensed phases
    of compounds in question are not thermodynamically equilibrium
    states. For example, molecular ethylene $C_2H_4$, being a simple compound,
    is not the ground state of the system; in particular,
    polyethylene $C_{2n}H_{4n}$ is a more low-laying energy modification \cite{8}.
    Moreover, on the given composition it is the mixture of solid carbon
    (graphite) and methane ($CH_4$) in the ratio of 1 : 1 that will
    constitute an equilibrium state. Thus the only present
    substances in the equilibrium concentration diagram C - H
    at ambient pressure in a wide temperature range are pure hydrogen,
    graphite, and a single compound, methane, while other numerous
    molecular phases like ethylene, acetylene, benzene are metastable
    with regard to the transition to the methane-graphite mixture in
    the proper ratio. That this assertion is true can be clearly seen
    from the data on the bonding energies of molecules \cite{9,10}:
    for $H_2 - 430 kJ/mole, CH_4 - 1642 kJ/mole, C_2H_4 - 2225 kJ/mole,
    CH_2 - 753 kJ/mole, C_2H_2 - 1626 kJ/mole$, cohesive energy for
    graphite - 712 kJ/mole, the energies of the formation of
    condensed phases from molecules $CH_2, C_2H_4, C_2H_2, C_6H_6,
    CH_4$
    and others do not exceed tens of kJ/mole. At extremely high
    temperatures and low pressure in a gas medium the molecular
    states at the intermediate compositions match the Gibbs' free
    energy minimum (Fig.3). The situation changes with the decrease
    in temperature and carbon condensation: most of hydrocarbon
    molecular phases are to be found only within the local minimums
    and are separated from the equilibrium state, the graphite-methane
    mixture, by a high energy barrier (Fig.3). In a similar way,
    the only present compounds in the equilibrium concentration
    diagram in the system C-O-H, besides pure components,
    are $H_2O$, $CH_4$ and $CO_2$, while other compounds, including,
    for example, $CO$, alcohols or glycerin, are metastable phases
    (although they melt and boil reversibly). The fact that most
    of hydrocarbons at ambient pressure are not in the equilibrium state
    has been known to chemists for quite a time (see, for example,
    \cite{11} and references therein). On the other hand, the effective
    range of pressures $P \leq 1 GPa$ and temperatures $T \leq500 K$, employed in
    chemistry, corresponds to the quasi-equilibrium region in
    the P,T transformations diagram (see Fig.2), where a researcher
    can "forget" about the metastable state of phases under
    investigation. In resent years, however, the substances in question
    have been studied in a wider P,T range ($P>10 GPa$) beyond the
    quasi-equilibrium region \cite{12,13,14,15,16,17,18}. In the process,
    the studies have often involved P,T phase diagrams of different
    molecular compounds of the same composition, for example,
    acetylene $C_2H_2$ and benzene $C_6H_6$ \cite{12,13}, the phases of both
    substances being metastable. In the quasi-equilibrium region
    of pressures and temperatures (Fig.2) the properties, structure,
    and aggregate state of such substances changes reversibly; however,
    at sufficiently high pressure irreversible transformations to more
    low-laying energy states, including polymerization, begin. In particular,
    acetylene and benzene irreversibly transform to the polymeric phases,
    and then to amorphous diamond-like carbon saturated by hydrogen a-C:H
    \cite{13,14,16,17}. In so doing, the non-equilibrium part of
    the kinetic transformations diagram is realized (Fig.2).

\begin{figure}
\includegraphics[width=7cm]{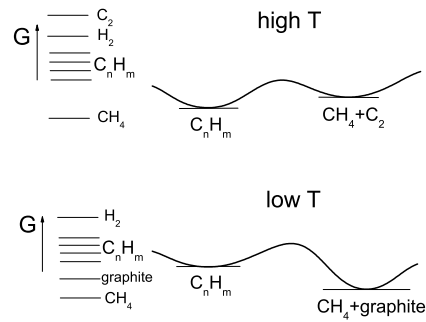}%
\caption{\label{fig:fig3} Gibbs' free energy for carbon, hydrogen
and hydrocarbons at high temperatures (gas phases) and low
temperatures (condensed phases). At low temperatures the
equilibrium state is the methane-graphite mixture, and all other
hydrocarbons are in metastable state. }
\end{figure}

    Metastable substances cannot exist for a significant time
    at temperatures above the melting temperature of a stable phase
    \cite{8}. In particular, high-pressure metallic modifications
    of hydrogen and oxygen, and a polymeric modification of nitrogen
    at ambient pressure can potentially exist in the metastable state
    only at ultra-low temperatures below the melting temperatures
    of stable molecular phases. However, as was noted in \cite{8},
    if the initial molecular phase is metastable by itself, then more
    low-laying energy modifications obtained from it can of course
    exist in the metastable state at temperatures above the melting t
    emperature of the molecular phase in question. The work \cite{8}
    has offered a bright example of such metastable state, namely,
    polyethylene, whose temperature stability far exceeds the one
    for the condensed phases of molecular ethylene and is fundamentally
    limited by a methane-graphite liquidus line. We again point out
    to the irreversibility of the transformation under pressure
    of metastable molecular phases to polymeric modifications:
    on annealing, such molecular phases never transform back to the
    molecular states.

{\bf 3}. In addition to hydrocarbons and alcohols, which are
nevertheless organic substances, there are, of course, other
numerous examples of metastable "kinetic" phases of simple
inorganic substances. Thus, in the system N-H at normal pressure
there is only one equilibrium phase, ammonia NH3, while NH2, N2H2,
N2H4, HN3 and other molecular phases are metastable in relation to
the decomposition to ammonia and molecular nitrogen (the
corresponding molecule bonding energies are: $N_2 - 942 kJ/mole,
NH_2 - 710 kJ/mole, NH_3 - 1158 kJ/mole, N_2H_2 - 1154 kJ/mole,
N_2H_4 - 1696 kJ/mole, HN_3 - 1328 kJ/mole$ \cite{9}). In the
system N-O, metastable molecular phases are $N_2O, NO_2, N_2O_3,
N_2O_4, N_2O_5$ (the corresponding molecule bonding energies are:
$N_2O - 1103 kJ/mole, NO_2 - 927 kJ/mole, N_2O_3 - 1590 kJ/mole,
N_2O_4 - 1908 kJ/mole, N_2O_5 - 2153 kJ/mole$ \cite{9}). In the
system C-N, also metastable in relation to the decomposition to
solid carbon and molecular nitrogen are molecular phases $CN, NCN,
CNC, CCN, C_2N_2, C_3N_4$ \cite{10}. Metastable phases are some
alkali, acids and many molecular compounds with the participation
of the phosphorous and sulfur atoms.

Let us discuss separately such system as molecular carbon oxide.
In recent years, the phase transformations under pressure in this
system have been intensively studied \cite{18,19,20} and
references therein], but the obtained experimental data have been
often interpreted erroneously. In the equilibrium concentration
diagram in the carbon-oxygen system at normal pressure there is
only one compound present, $CO_2$. (The $CO_2$ formation energy is
$1598 kJ/mole, CO - 1071 kJ/mole$ \cite{10}). The $CO$ molecular
condensed phases are metastable with relation to the decay into
the mixture of solid carbon (graphite) and $CO_2$; this
distinguishes the $CO$ compound from its isoelectronic analog,
$N_2$, for which the molecular phases at ambient pressure are
equilibrium. The examination of the $CO$ phase diagram at high
pressures revealed the transformation to a $CO$ polymerized
modification, erroneously taken to be a high-pressure phase of the
molecular phase \cite{18,19}. In actuality the polymerization of
the CO metastable molecular phase at P about 4-5 GPa constitutes a
non-equilibrium kinetic transformation to a more low-laying energy
state. According to computer simulation data the CO polymerization
is followed by a considerable heat emission \cite{20}. The $CO$
polymerized phase is metastable at normal pressure up to high
temperatures (400-700 K) \cite{19}, which is much higher than the
melting temperature of the $CO$ molecular crystal. As with
polyethylene, it is hardly surprising that polymerized $CO$
demonstrates a high thermal stability, since a polymerized phase
is not a high-pressure phase of a molecular modification but is a
lower energy state. If heating the $CO$ polymerized phase at
normal pressure, a significant amount of heat is being released,
which was mistakenly ascribed by the authors \cite{19} to the
energy stored up in the polymerized phase (and being excess
compared to the molecular phase) at the expense of the PV-terms in
the Gibbs' free energy. At the same time, a simple calculation of
the difference of the PV-terms for the polymerized phase as
against the molecular modification for the transition pressure
about 4 GPa gives the value about 0.5-1 kJ/g, which is an order
less than the observed values. In reality, the observed energy
release is apparently associated with the transition of the $CO$
polymerized phase to an equilibrium state of the mixture of $CO_2$
and solid carbon (these are precisely the products found after the
annealing of polymer $CO$). Note that the pattern of
non-equilibrium transformations under pressure of the $CO$
molecular phase can be more complicated; in particular, under
pressure a disproportionation is possible of the CO compound into
$C_2O_3$ and $C_3O_2$ modifications which, in turn, are
polymerized \cite{19,20,21}. In any event the polymerization of
$CO$ under pressure is a non-equilibrium transformation to a more
low-laying energy state similar to the polymerization of acetylene
$C_2H_2$ and benzene $C_6H_6$ under pressure.

    The reason why there are so many long-lived metastable
    "kinetic" phases among the light element compounds is a small
    size of the corresponding molecules. The distance between
    the atoms of light elements inside a molecule is noticeably
    less than the distance between the atoms of different molecules.
    The molecules are bonded by the van der Waals forces which are
    two orders weaker than the interaction inside the molecules.
    At the distances that correspond to the minimum of intermolecular
    interaction potential, the molecules appear quite small,
    and the separation of the substances into molecules
    is determined quite well. It is easy to understand that molecular
    substances subjected to pressure within a particular range reach
    the boundaries of the quasi-equilibrium P,T region providing the
    distance between molecules becomes comparable to the distances
    between the atoms inside a molecule, which involves a several
    times decrease in the volume of substances. Characteristic values
    of the bulk modulus of substances with molecular interaction amount
    to units of GPa, hence, characteristic pressures of irreversibility
    are of the order of 10 GPa.

Thus, the majority of molecular phases of the light element
compounds are not stable for a given chemical composition, while
in the equilibrium concentration diagrams at low pressures and
temperatures there are only several phases, such as $C, O_2, H_2,
N_2, H_2O, CO_2, CH_4, NH_3$. As we noted above, high pressures
can be a contributory factor for lowering the energy barrier for
the transformation of the metastable molecular phases to more
equilibrium states. At the same time, at sufficiently high
pressures the equilibrium concentration diagrams themselves can
undergo changes because of the different contributions of PV-terms
for the phases of a variable density; in particular, a number of
unsaturated hydrocarbons can become more advantageous in terms of
energy than the mixture of the condensed phases of methane and
carbon. As a result, sufficiently complex polymerized phases in
the light element systems may emerge at compression not only as
intermediate in energy states during the transformation of the
non-equilibrium molecular phases, but also as ground, most stable
at high pressure modifications. For example, at certain pressures
the formation of higher hydrocarbons becomes beneficial, which is
of great importance for the abiogenous petroleum synthesis
(petroleum at normal pressure is, of course, composed of the
metastable phases of hydrocarbons) \cite{11,22}.

\begin{figure}
\includegraphics[width=7cm]{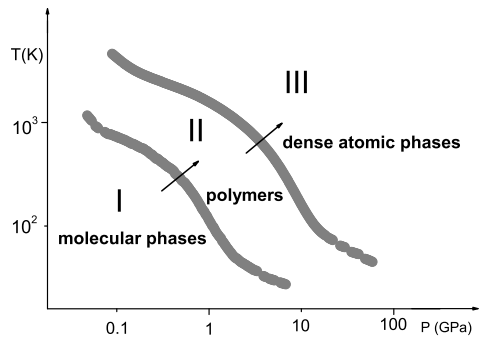}%
\caption{\label{fig:fig4} General kind of a transitional P,T-phase
diagram of molecular low Z- elements compounds. In most cases the
transitions between the phases from Zone I to Zone II and from
Zone II to Zone III are non-equilibrium "kinetic" transformations.
}
\end{figure}

    To sum up, all P,T-diagrams of the light element molecular
    compounds can be arbitrarily divided into three zones (Fig.4).
    Zone I includes moderate pressures and temperatures (1-10 GPa, 10-1000 K),
    at which the Gibbs' free energy of the phases is almost completely
    determined by the bonding energy of atoms in molecules. Most of
    molecular substances in this zone are in the metastable state.
    Zone III involves ultrahigh pressures (20-50 GPa), at which the
    Gibbs'
    free energy of the majority of the phases is largely determined
    by a PV-term. The substances in this zone are in the stable state,
    as a rule, and represent the mixture of simple solid equilibrium
    modifications (diamond, $CO_2, H_2O$, etc.). Specifically, at megabar
    pressures 20-50 GPa and moderate temperatures the PV-contribution
    to the Gibbs' free energy facilitates the transition of all
    unsaturated hydrocarbons to a solid hydrogen solution in diamond;
    at higher temperatures it brings about the formation of a
    diamond-hydrogen mixture. As was pointed out above,
    the formation of amorphous diamond-like carbon saturated with
    hydrogen a-C:H, indeed, was observed for many hydrocarbons at
    ultrahigh pressures \cite{13,14,16,17}. Zone II corresponds
    to intermediate pressures 1-20 GPa and temperatures 200-1500 K,
    at which the formation of a large number of both "kinetic"
    and equilibrium, at a given pressure, polymerized phases is
    possible. The contributions to the Gibbs' free energy from
    intra-molecular, intermolecular interaction and PV-contribution
    for these phases are comparable in magnitude. One should only
    remember that the modifications in question, as a rule, are not
    the high-pressure phases of molecular substances, since the majority
    of molecular substances at ambient pressure are metastable themselves.
    The high-pressure experiments, in this case, suggest what intermediate
    in energy polymeric phases of a substance can exist. Evidently,
    many of these uncommon polymeric phases can be synthesized through
    "chemical" techniques at normal pressure like well-known examples
    of polyethylene and polyacetylene.

\begin{acknowledgments}
The author is grateful to A.G. Lyapin, S.V. Popova, N.A.
Bendeliani, S.M. Stishov, Yu.A. Lozovik and V.N. Ryzhov for useful
discussions. The work was supported by the Russian Foundation for
Basic Research, project nos. 05-02-16596 and 04-02-16308, the
Program of the Presidium of the Russian Academy of Sciences, and
Russian Science Support Foundation.
\end{acknowledgments}


\end{document}